\documentclass[aps,prd,twocolumn,showpacs,groupedaddress,nofootinbib,superscriptaddress,10pt]{revtex4-1}

\usepackage{newtxtext}

\usepackage[T1]{fontenc}
\usepackage{lmodern}

\usepackage[dvipsnames,x11names]{xcolor}

\usepackage[colorlinks=true,linkcolor=Blue,citecolor=ForestGreen,urlcolor=NavyBlue]{hyperref}

\usepackage{graphicx}
\usepackage[sort&compress]{natbib}
\usepackage{lipsum}
\usepackage{morefloats}
\usepackage[pdf]{pstricks}
\usepackage{subfigure}
\usepackage{amsmath}
\usepackage{amssymb}
\usepackage{amsfonts}
\usepackage{rotating}
\usepackage{cancel}
\usepackage{mathtools}
\usepackage{bbm}
\usepackage{dsfont}
\usepackage{bbold}
\usepackage{multirow}
\usepackage{ulem}
\usepackage{physics}
\newcommand{\mev}{\textrm{ MeV}}

\newcommand{\GXNU}{Department of Physics, Guangxi Normal University, Guilin 541004, China}
\newcommand{\GXZD}{Guangxi Key Laboratory of Nuclear Physics and Technology, Guangxi Normal University, Guilin 541004, China}
\newcommand{\IFIC}{Departamento de F\'{\i}sica Te\'orica and IFIC, Centro Mixto Universidad de
Valencia-CSIC Institutos de Investigaci\'on de Paterna, Aptdo.22085,
46071 Valencia, Spain}

\begin{document}

\frenchspacing

\title{\boldmath $D^+ \to K_s^0 \pi^+ \eta$ reaction and $a_0(980)^+$}

\author{Natsumi Ikeno}
 \email{ikeno@tottori-u.ac.jp}
 \affiliation{Department of Agricultural, Life and Environmental Sciences, Tottori University, Tottori 680-8551, Japan}

\author{Jorgivan M. Dias}%
\email{jorgivan.mdias@itp.ac.cn}
\affiliation{CAS Key Laboratory of Theoretical Physics, Institute of Theoretical Physics, Chinese Academy of Sciences, Beijing 100190, China}%

\author{Wei-Hong Liang}%
\email{liangwh@gxnu.edu.cn}
\affiliation{\GXNU}%
\affiliation{\GXZD}%

\author{Eulogio Oset}%
\email{Eulogio.Oset@ific.uv.es}
\affiliation{\IFIC}%
\affiliation{\GXNU}%

\begin{abstract}
We study the $D^+ \to \bar K^0 \pi^+ \eta$ reaction where the $a_0(980)$ excitation plays a dominant role. We consider mechanisms of external and internal emission at the quark level, hadronize the $q \bar q$ components into two mesons and allow these mesons to undergo final state interaction where the $a_0(980)$ state is generated. While the $a_0(980)$ production is the dominant term, we also find other terms in the reaction that interfere with this production mode and, through interference with it, lead to a shape of the $a_0(980)$ significantly different from the one observed in other experiments, with an apparently much larger width.

\end{abstract}

\maketitle

\section{Introduction}
The $D^+ \to K_s^0 \pi^+ \eta$ reaction measured by the BESIII collaboration in Ref.~\cite{BESIII:2020pxp}, and more recently in Ref.~\cite{BESIII:2023htx} with more precision and an amplitude analysis, has turned into an ideal reaction to isolate the $a_0(980)$ contribution.
The reaction is actually $D^+ \to \bar K^0 \pi^+ \eta$, and the $\bar K ^0$ is observed as a $K_s^0$ state.
It might look at a simple sight that this reaction is just a copy of the $D^0 \to K^- \pi^+ \eta$ reaction measured by the Belle collaboration \cite{Belle:2020fbd}.
Indeed, one goes from the first reaction to the second by changing a $\bar d \to \bar u$ quark. Yet, the differences are striking as we shall see in the present work, comparing it with the theoretical study of the $D^0 \to K^- \pi^+ \eta$ reaction in Ref.~\cite{Toledo:2020zxj}.
The key point to see these striking differences is the fact that the $K^- \pi^+$ in the $D^0 \to K^- \pi^+ \eta$ reaction can come from $\bar K^{*0}$ excitation, but $\bar K^0 \pi^+$ with positive charge and an $s$ quark cannot come from $\bar K^{*0}, K^{*-}$.
Hence, there is no $\bar K^*$ contribution in the $D^+ \to \bar K^0 \pi^+ \eta$ reaction which makes cleaner the $a_0(980)$ production as seen in the experiment \cite{BESIII:2023htx}.

The differences in the mass distributions in the $D^0 \to K^- \pi^+ \eta$ and $D^+ \to \bar K^0 \pi^+ \eta$ reactions are striking.
Indeed, in the $D^0 \to K^- \pi^+ \eta$ reaction the $K^- \pi^+$ mass distribution is dominated by the $\bar K^{*0}$ contribution, with a sharp peak at the $\bar K^{*0}$ mass, while in the $D^+ \to \bar K^0 \pi^+ \eta$ reaction the $\bar K^0 \pi^+$ mass distribution is rather structureless.
This is not all, because the $\pi^+ \eta$ mass distribution in the $D^0 \to K^- \pi^+ \eta$ reaction, shows indeed a peak for the $a_0(980)$ production but has large strength at low and high invariant masses which are replicas of the $\bar K^{*0}$ resonance in the $K^- \pi^+$ channel.
Even more striking is the shape of the $K^- \eta$ mass distribution which has a double hump shape created again by the presence of the $\bar K^{*0}$ resonance in the $K^- \pi^+$ channel.
 By contrast, the $\pi^+ \eta$ mass distribution in the $D^+ \to \bar K^0 \pi^+ \eta$ reaction has a mass distribution where the $a_0(980)$ dominates the spectrum and the $\bar K^0 \eta$ mass distribution has also very different shape.

It is worth mentioning that the related $D^0 \to K_s^0 \pi^+ \pi^-, \bar K_s^0 \pi^0 \eta$ reactions were studied prior to the experiments, paying attention to the $\pi^+\pi^-$ and $\pi^0\eta$ mass distributions, predicting that a clear signal of the $a_0(980)$ should be seen in these experiments~\cite{Xie:2014tma}.

Our purpose in the present work is to try to understand the spectrum from the perspective that the $a_0(980)$ is a dynamically generated state from the interaction of the $\pi \eta, K\bar K$ channels, which is well described by the chiral unitary approach \cite{Oller:1997ti,Kaiser:1998fi,Markushin:2000fa,Nieves:1998hp}.
For this we shall investigate the reaction mechanism of external and internal emission \cite{Chau:1982da} at the quark level, and see how the hadronization of $q\bar q$ pairs can lead to the production of the final state.

\section{Formalism}
\label{sec:formula}

We study the $D^+ \to \bar K^0 \pi^+ \eta$ reaction and look at the mechanisms for external and internal emission.

\subsection{External emission}
\label{subsec:Ee}

\begin{figure*}[bt!]
  \begin{center}
  \includegraphics[scale=0.6]{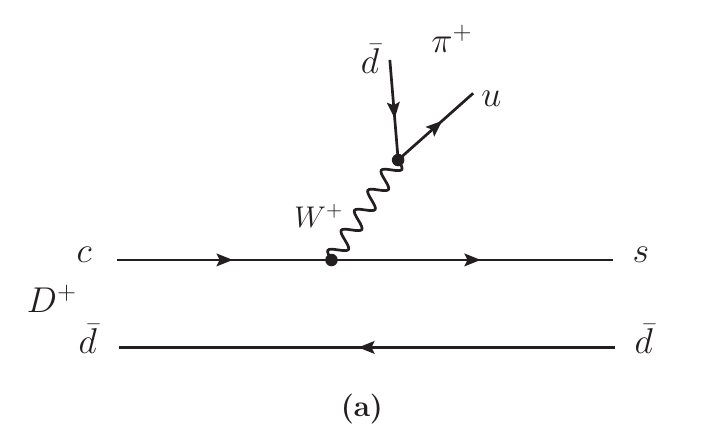}~~~~
  \includegraphics[scale=0.6]{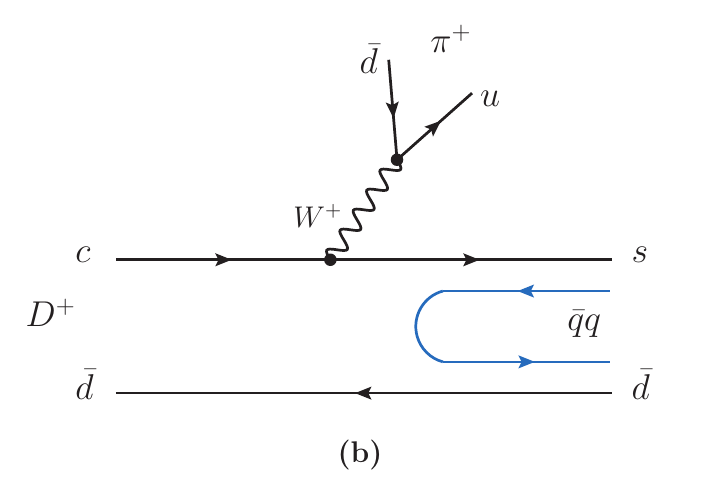}
  \end{center}
  \vspace{-0.5cm}
  \caption{Mechanism with external emission at the quark level. (a) $D^+ \to \pi^+ s \bar d$; (b) hadronization of the $s\bar d$ component adding a $\bar qq$ pair producing two mesons.}
  \label{fig:Fig1}
  \end{figure*}

In Fig. \ref{fig:Fig1}(a)  
we show the mechanism of external emission, Cabibbo favoured, and in Fig. \ref{fig:Fig1}(b) the $s\bar d$ pair is hadronized in the following way
\begin{equation}
  s\bar d \to \sum_i s\;\bar q_i q_i\; \bar d,
\end{equation}
which is easily interpreted writing the $q_i \bar q_j$ matrix in terms of pseudoscalar mesons, $\mathcal{P}_{ij}$, which in the standard $\eta-\eta'$ mixing of Ref.~\cite{Bramon:1992kr} is given by

\begin{equation}\label{eq:Pmatrix}
  \mathcal{P} =
   \left(
   \begin{array}{cccc}
   \frac{1}{\sqrt{2}}\pi^0 + \frac{1}{\sqrt{3}}\eta  & \pi^+ & K^+  \\[2mm]
   \pi^- & -\frac{1}{\sqrt{2}}\pi^0 + \frac{1}{\sqrt{3}}\eta & K^0  \\[2mm]
   K^- & \bar{K}^0 & ~-\frac{1}{\sqrt{3}}\eta ~  \\
   \end{array}
   \right).
\end{equation}
Hence,
\begin{equation}\label{eq:sd}
  s\bar d \to \sum_i \;\mathcal{P}_{3i} \; \mathcal{P}_{i2}=\left( \mathcal{P}^2\right)_{32}=K^-\pi^+-\bar K^0 \,\dfrac{\pi^0}{\sqrt{2}},
\end{equation}
where the $\bar K^0 \eta$ component has cancelled.

The combination of Eq.~\eqref{eq:sd}, together with the $\pi^+$ of Fig.~\ref{fig:Fig1}, does not lead to the desired final state $\bar K^0 \pi^+ \eta$.
However, through rescattering the $K^-\pi^+$ and $\bar K^0 \pi^0$ could lead to $\bar K^0 \eta$.
Yet, the threshold of $\bar K^0 \eta$ is $1040\mev$, which is about $300\mev$ above the peak of the $K^*_0(700)$ (the kappa), where the $K^- \pi^+ \to \bar K^0 \eta$ amplitude has a reduced strength, and because of that we shall disregard this contribution, as it was also done in Ref.~\cite{Toledo:2020zxj}.

\begin{figure}[bt!]
\begin{center}
\includegraphics[scale=0.63]{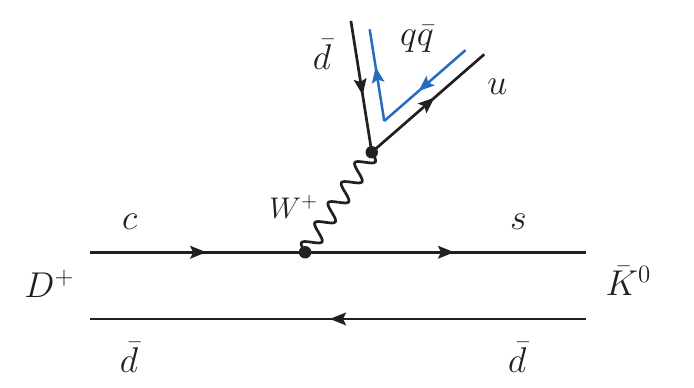}
\end{center}
\vspace{-0.5cm}
\caption{Hadronization of the $\bar d u$ component in external emission.}
\label{fig:Fig2}
\end{figure}

Next, with the same mechanism of Fig.~\ref{fig:Fig1}(a), we look at the hadronization of the $\bar d u$ component, as shown in Fig.~\ref{fig:Fig2}.
We have now
\begin{eqnarray}\label{eq:ud}
  u\bar d &\to  & \sum_i u \;\bar q_i q_i \;\bar d =\sum_i \;\mathcal{P}_{1i} \; \mathcal{P}_{i2}=\left( \mathcal{P}^2\right)_{12} \nonumber\\
&=&\dfrac{2}{\sqrt{3}}\, \eta \pi^+ + K^+ \bar K^0,
\end{eqnarray}
where now the $\pi^0\pi^+$ component has cancelled.

Hence we have now the combination
\begin{equation}\label{eq:H1}
  H=\left( \frac{2}{\sqrt{3}}\, \eta \pi^+ + K^+ \bar K^0 \right) \, \bar K^0.
\end{equation}
However, there is a subtlety here concerning the $K^+ \bar K^0$ production (see discussion in page 3 of Ref.~\cite{Ikeno:2021kzf}) because for dynamical reasons the $WK^+\bar K^0$ vertex goes as the difference of energies of $K^+\bar K^0$ which vanishes in the average.
However, due to the different masses of $\eta\pi^+$ this cancellation does not occur, and consequently we keep the $\eta \pi^+$ term and disregard the $K^+\bar K^0$ one.

\begin{figure*}[tb!]
  \begin{center}
  \includegraphics[scale=0.63]{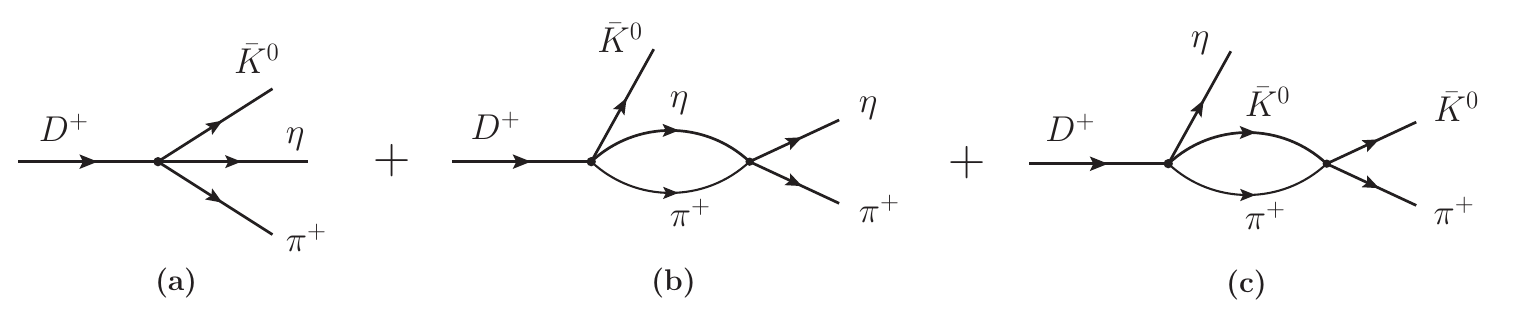}
  \end{center}
  \vspace{-0.5cm}
  \caption{Diagrams for $D^+ \to \bar K^0 \pi^+ \eta$ coming from external emission: (a) tree level; (b) $\eta \pi^+$ rescattering; (c) $\bar K^0 \pi^+$ rescattering.}
  \label{fig:Fig3}
  \end{figure*}

We can see that the term $H$ in Eq.~\eqref{eq:H1} already contains the $\bar K^0 \pi^+ \eta$ state at the tree level.
In addition we can have rescattering of the $\eta \pi^+$ and also of the $\bar K^0 \pi^+$. Once again, we disregard the $\eta \bar K^0$ rescattering for the reasons given before.
Then, the picture that we have for $\bar K^0 \pi^+ \eta$ production is shown in Fig.~\ref{fig:Fig3}, including tree level and rescattering.

Analytically we have for these terms originating from external emission
\begin{eqnarray}\label{eq:tee}
    t^{(ee)}&=& \mathcal{C} \left\{h_{\eta \pi^+ \bar K^0}  \right.\nonumber\\[2mm] 
& & + \left. h_{\eta \pi^+ \bar K^0} \left[ G_{\eta \pi^+} (M_{\rm inv}(\eta \pi^+)) \cdot t_{\eta \pi^+, \, \eta \pi^+} (M_{\rm inv}(\eta \pi^+)) \right. \right.\nonumber\\[2mm]
    && +\left. \left. G_{\bar K^0 \pi^+} (M_{\rm inv} (\bar K^0 \pi^+)) \cdot t_{\bar K^0 \pi^+, \, \bar K^0 \pi^+}(M_{\rm inv}(\bar K^0 \pi^+)) \right] \right\}, \nonumber\\
\end{eqnarray}
where $h_{\eta \pi^+ \bar K^0}$ is the weight of the $\eta \pi^+ \bar K^0$ component in Eq.~\eqref{eq:H1},
\begin{equation}
  h_{\eta \pi^+ \bar K^0}= \dfrac{2}{\sqrt{3}},
\end{equation}
and $G_{i}, t_{ij}$ are the loop functions of two mesons and the scattering matrices for the transition of channel $i$ to channel $j$, taken as in Ref.~\cite{Toledo:2020zxj}.
$\mathcal{C}$ in Eq.~\eqref{eq:tee} is a global constant that will be used to get the normalization of the data.

\subsection{Internal emission}
\label{subsec:Ie}

\begin{figure*}[bt!]
  \begin{center}
  \includegraphics[scale=0.6]{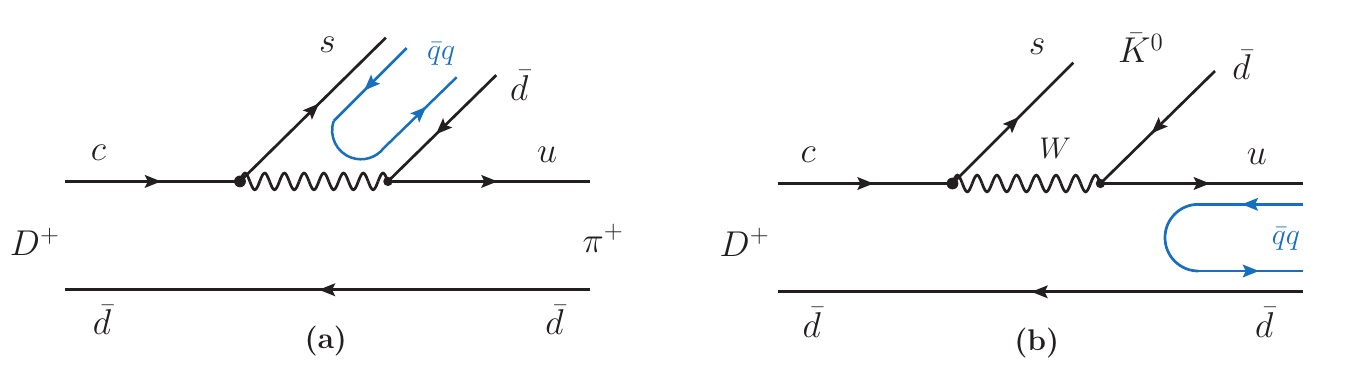}
  \end{center}
  \vspace{-0.5cm}
  \caption{Diagrams for internal emission: (a) hadronization of the $s\bar d$ pair; (b) hadronization of the $u\bar d$ pair.}
  \label{fig:Fig4}
  \end{figure*}

\begin{figure*}[tb!]
  \begin{center}
  \includegraphics[scale=0.6]{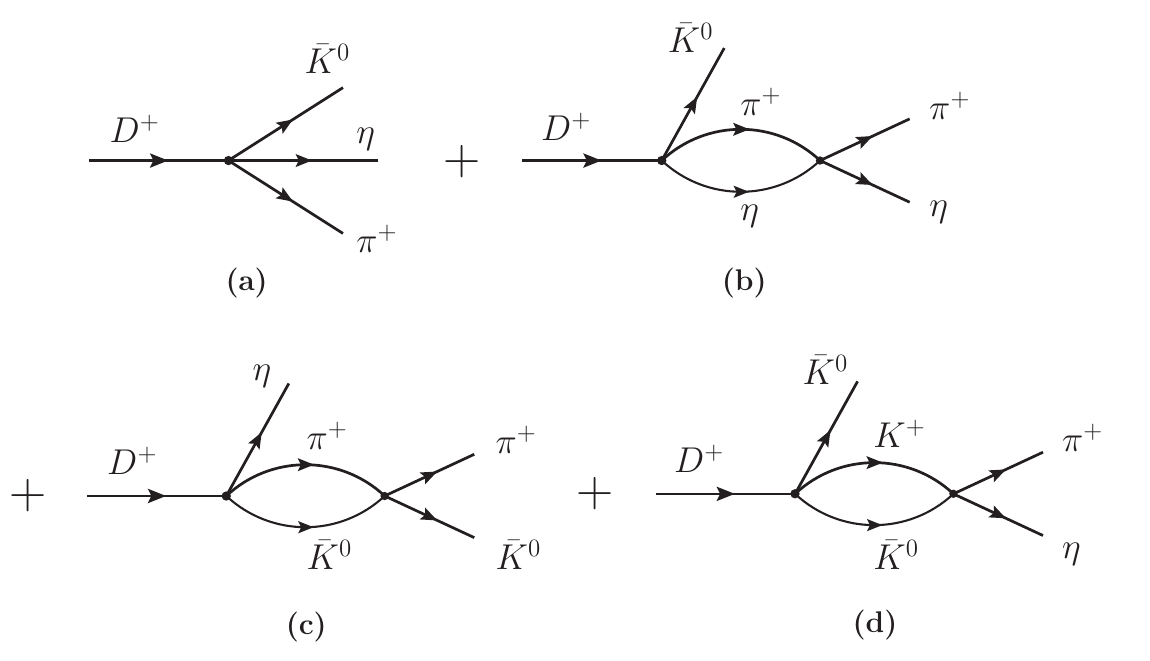}
  \end{center}
  \vspace{-0.5cm}
  \caption{Diagrams for $\eta \pi^+ \bar K^0$ production coming from internal emission: (a) tree level; (b) $\pi^+ \eta$ rescattering; (c) $\pi^+ \bar K^0$ rescattering; (d) $K^+ \bar K^0$ rescattering.}
  \label{fig:Fig5}
  \end{figure*}

For internal emission we have the diagrams of Fig.~\ref{fig:Fig4}, which account for the hadronization of quark pairs.
Following the path of subsection \ref{subsec:Ee}, we have now for the hadronization
\begin{eqnarray}\label{eq:sd2}
  s\bar d &\to& \sum_i s \;\bar q_i q_i \;\bar d =\sum_i \;\mathcal{P}_{3i} \; \mathcal{P}_{i2}=\left( \mathcal{P}^2\right)_{32} \nonumber\\
&=&K^- \pi^+ -\dfrac{1}{\sqrt{2}}\, \pi^0 \bar K^0,
\end{eqnarray}
where the $\bar K^0 \eta$ terms have cancelled, and
\begin{eqnarray}\label{eq:ud2}
  u\bar d &\to& \sum_i u \;\bar q_i q_i \;\bar d =\sum_i \;\mathcal{P}_{1i} \; \mathcal{P}_{i2}=\left( \mathcal{P}^2\right)_{12} \nonumber\\
&=& \dfrac{2}{\sqrt{3}}\, \eta \pi^+ + K^+ \bar K^0,
\end{eqnarray}
where the $\pi^+ \pi^0$ terms have cancelled.
Summing the two terms we have, including the $\pi^+$ in Fig.~\ref{fig:Fig4}(a) and $\bar K^0$ in Fig.~\ref{fig:Fig4}(b),
\begin{equation}\label{eq:H'}
  H'=K^- \pi^+ \pi^+ -\dfrac{1}{\sqrt{2}}\; \pi^0 \pi^+ \bar K^0 + \dfrac{2}{\sqrt{3}}\; \eta \pi^+ \bar K^0 + K^+ \bar K^0 \bar K^0.
\end{equation}
Once again we have a tree level $\eta \pi^+ \bar K^0$ production and the other terms can lead to this final state through rescattering, as shown in Fig.~\ref{fig:Fig5}.

  In Fig.~\ref{fig:Fig5} we have disregarded the possible rescattering of $K^- \pi^+ \to \bar K^0 \eta$, $\bar K^0 \pi^0 \to \bar K^0 \eta$, $\bar K^0 \eta \to \bar K^0 \eta$ for the same reasons as in the former subsection.
  We have checked numerically that these terms are, indeed, negligible.

  Analytically we have for the diagrams of Fig.~\ref{fig:Fig5} corresponding to internal emission
\begin{eqnarray}\label{eq:tie}
    t^{(ie)}&=& \beta \mathcal{C}  \left\{\bar h_{\eta \pi^+ \bar K^0}  \right.\nonumber\\[2mm]
&+& \left. \bar h_{\eta \pi^+ \bar K^0} \left[ G_{\eta \pi^+} (M_{\rm inv}(\eta \pi^+)) \cdot t_{\eta \pi^+, \, \eta \pi^+} (M_{\rm inv}(\eta \pi^+)) \right. \right.\nonumber\\[2mm]
 &+& \left. G_{\pi^+ \bar K^0} (M_{\rm inv} (\pi^+ \bar K^0)) \cdot t_{\pi^+ \bar K^0, \, \pi^+ \bar K^0}(M_{\rm inv}(\pi^+ \bar K^0)) \right] \nonumber\\[2mm]
    &+& 2 \left. \bar h_{K^+ \bar K^0 \bar K^0} \; G_{K^+ \bar K^0}(M_{\rm inv}(\pi^+ \eta)) \right.\nonumber\\[2mm]
& & \left.\ \  \cdot t_{K^+ \bar K^0, \pi^+ \eta} (M_{\rm inv}(\pi^+ \eta)) \right\}, 
 \end{eqnarray}
where the weights $\bar h_i$ are now
\begin{equation}
  \bar h_{\eta \pi^+ \bar K^0}= \dfrac{2}{\sqrt{3}};~~~
  \bar h_{K^+ \bar K^0 \bar K^0}=1,
\end{equation}
and the factor $2$ multiplying $\bar h_{K^+ \bar K^0 \bar K^0}$ in Eq.~\eqref{eq:tie} accounts for the two $\bar K^0$ identical particles.
In Eq.~\eqref{eq:tie} we have put a weight $\beta \mathcal{C} $, conserving the global factor $\mathcal{C}$.
The $\beta$ should be a factor of the order of $\frac{1}{N_c}$ ($N_c$ number of colors) which one expects from the ratio of terms from internal emission to those of external emission.

The total amplitude for the process is now given by
\begin{equation}\label{eq:ttotal}
  t=t^{(ee)} + t^{(ie)}.
\end{equation}

\subsection{$K^*_0(1430)$ contribution}
\label{subsec:K0}
\begin{figure}[tb!]
  \begin{center}
  \includegraphics[scale=0.66]{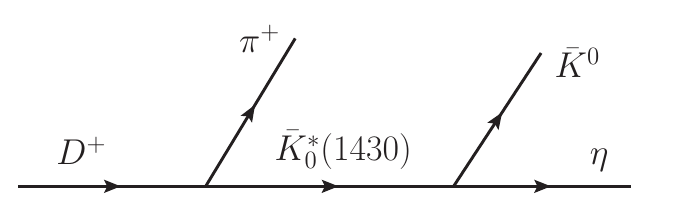}
  \end{center}
  \vspace{-0.5cm}
  \caption{Diagrammatic representation of $D^+ \to \pi^+ \bar K^*_0(1430) \to \pi^+ \bar K^0 \eta$.}
  \label{fig:Fig6}
  \end{figure}

In Ref.~\cite{BESIII:2023htx} it was found that the scalar $K^*_0(1430)$ [$I\, (J^P)=\frac{1}{2}\, (0^+)$] state contributed to the process and shows up in the $K\eta$ mass distribution.
The process can proceed as shown in Fig.~\ref{fig:Fig6}.
We take into account this contribution phenomenologically by means of the amplitude
\begin{equation}\label{eq:K0star}
  t^*= \mathcal{D} \, e^{i \phi}\; \dfrac{M_D^2}{s_{13}-M^2_{K^*_0}+i M_{K^*_0}\, \Gamma_{K^*_0}},
\end{equation}
where $\mathcal{D}$ and $\phi$ will be chosen as free parameters and $s_{13}=(p_{\bar K^0}+p_\eta)^2$ taking the order of the particles $\bar K^0 (1), \pi^+ (2), \eta (3)$. The factor $M_D^2$ is put to have $\mathcal{D}$ dimensionless.
The final amplitude will now be
\begin{equation}\label{eq:ttotal2}
  t=t^{(ee)} + t^{(ie)} + t^*.
\end{equation}

Then in order to calculate the mass distributions, we use the PDG standard formula \cite{pdg2023}
\begin{equation}\label{eq:Gamm}
  \dfrac{d^2 \Gamma}{ds_{12}\; ds_{23}} = \dfrac{1}{(2\pi)^3}\; \dfrac{1}{32\, M^3_D}\; |t|^2.
\end{equation}
We can integrate over the limits of the PDG formula \cite{pdg2023} to get $d \Gamma / ds_{12}$ integrating over $s_{23}$.
Then, by cyclical permutations in the formulas we can get $d\Gamma / ds_{13}$, $d \Gamma / ds_{23}$ and compare with experiment \cite{BESIII:2023htx}.

\subsection{Scattering amplitudes}
\label{subsec:tij}
We need the amplitudes
\begin{equation*}
  t_{\eta \pi^+, \, \eta \pi^+}, ~~ t_{\bar K^0 \pi^+, \, \bar K^0 \pi^+}, ~~ t_{\bar K^0 K^+, \, \pi^+ \eta}.
\end{equation*}
We use the results of Ref.~\cite{Lin:2021isc} (Eq.~(A.4) of Ref.~\cite{Lin:2021isc}), which considers explicitly the $\eta-\eta'$ mixing, and for the channels $K^+ K^- (1), K^0 \bar K^0 (2), \pi^0 \eta (3)$ one finds the matrix elements of the potential
\begin{alignat}{2}\label{eq:Vij}
& V_{11}=-\dfrac{s}{2f^2},        &\quad \quad &V_{12}= -\dfrac{s}{4f^2}, \nonumber\\[1mm] 
& V_{13}= -\dfrac{3s-2m^2_K-m^2_\eta}{3\sqrt{6}f^2}, & \quad \quad & V_{22}=-\dfrac{s}{2f^2},        \nonumber\\[1mm]
& V_{23}= -\dfrac{3s-2m^2_K-m^2_\eta}{3\sqrt{6}f^2}, &\quad \quad &V_{33}= -\dfrac{2m^2_\pi}{3f^2}.
\end{alignat}
  %
By evaluating the coupled channel $T$ matrix
\begin{equation}\label{eq:BS}
  T= [1-VG]^{-1}\, V,
\end{equation}
we find, considering that $|\pi^+\rangle =-|11\rangle$ of isospin,
\begin{equation}\label{eq:t1}
  t_{\eta \pi^+, \, \eta \pi^+} = t_{\eta \pi^0, \, \eta \pi^0}.
\end{equation}
Then, taking into account that $K^+K^-$ in term of $|I,\, I_3\rangle$ is
\begin{equation}
  K^+K^- = - \left( \dfrac{1}{\sqrt{2}} | 10 \rangle
   + \dfrac{1}{\sqrt{2}} | 00 \rangle \right),
\end{equation}
we find that
\begin{equation}\label{eq:t2}
  t_{\bar K^0 K^+, \, \pi^+ \eta} = \sqrt{2}\; t_{K^+ K^-,\, \pi^0 \eta}.
\end{equation}

We also need the $K\pi \to K\pi$ amplitudes, which we take from the Appendix of Ref.~\cite{Toledo:2020zxj}, using the channels $\pi^- K^+ (1), \pi^0 K^0 (2), \eta K^0 (3)$.
Once again, using isospin coefficients and $C$ parity $C \bar K^0 \pi^+ = K^0 \pi^-$, we obtain 
\begin{equation}\label{eq:t3}
  t_{\bar K^0 \pi^+, \, \bar K^0 \pi^+} = \dfrac{2}{3}\, T_{22} + \dfrac{1}{3} \, T_{11} + \dfrac{2\sqrt{2}}{3} \, T_{12}.
\end{equation}
Note that this amplitude has $I=3/2$ and hence does not contain the $K^*_0(700)$.

\section{Results}
\begin{figure}[!tb]
\centering
\includegraphics[width=\columnwidth]{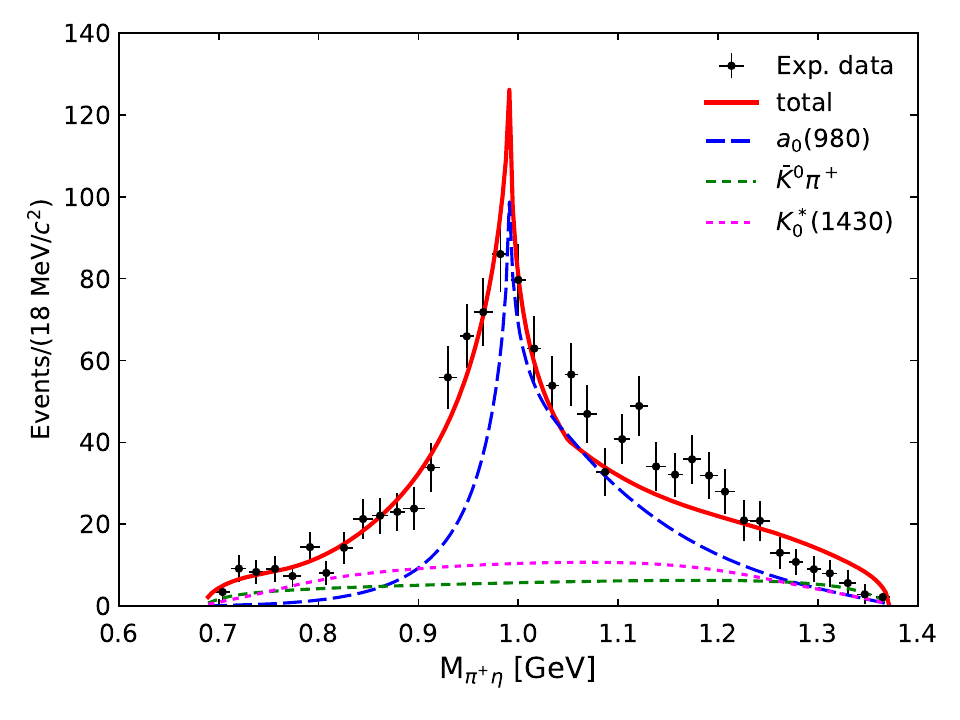}
\vspace{-0.5cm}
\caption{The $\pi^+ \eta$ mass distribution $d\Gamma/dM_{\pi^+ \eta}$.
The total theoretical result is shown in the red line.
The contributions of the $a_0(980)$, the $\bar K^0 \pi^+$ scattering terms, and the $K^*_0(1430)$ are shown in the blue, green, and magenta lines, respectively.
The experimental data shown by points are taken from Ref.~\cite{BESIII:2023htx}.
} 
\label{fig:dGdM_pieta}
 \end{figure}

\begin{figure}[!tbh]
\centering
\includegraphics[width=\columnwidth]{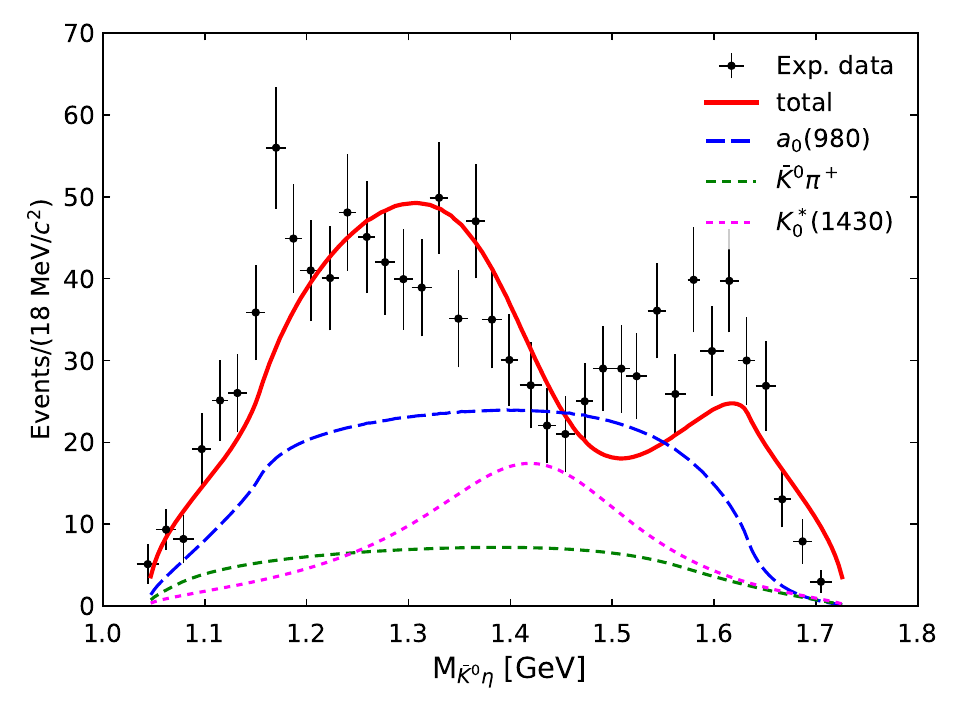}
\vspace{-0.5cm}
\caption{The $\bar K^0 \eta$ mass distribution $d\Gamma/dM_{\bar K^0 \eta}$.
The same as Fig.~\ref{fig:dGdM_pieta} but for the $\bar K^0 \eta$ mass distribution. } 
\label{fig:dGdM_keta}
\end{figure}

\begin{figure}[!tb]
\centering
\includegraphics[width=\columnwidth]{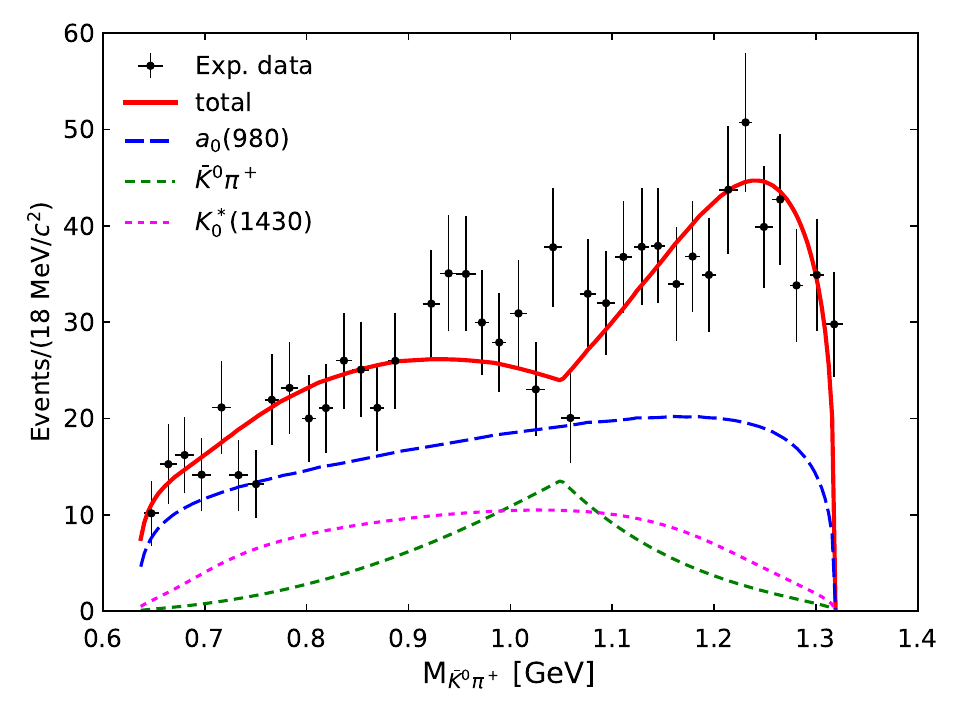}
 \vspace{-0.5cm}
\caption{The $\bar K^0 \pi^+$ mass distribution $d\Gamma/dM_{\bar K^0 \pi^+}$. The same as Fig.~\ref{fig:dGdM_pieta} but for the $\bar K^0 \pi^+$ mass distribution. } 
\label{fig:dGdM_kpi}
\end{figure}

We have four parameters, $\mathcal{C}$, $\beta$, $\mathcal{D}$, and $\exp(i \phi)$, in our theoretical model. 
To determine these parameters, we perform a best fit to the three mass distributions of $\bar K^0 \pi^+$, $\bar K^0 \eta$, and $\pi^+ \eta$ with the experimental data of Ref.~\cite{BESIII:2023htx}.
The parameter set obtained from the fit is shown in Table~\ref{tab:param}.
The parameter $\mathcal{C}$ provides a global normalization for external and internal emission. 
The parameter $\beta$ gives the relative weight of the internal emission mechanism to the external emission, while $\mathcal{D}$ quantifies the strength of the $K_0^*(1430)$ contribution. Additionally, the phase $\exp(i \phi)$ corresponds to the interference between the $K_0^*(1430)$ and other contributions, influencing the non-trivial shapes of the various mass distributions under consideration.
In Figs.~\ref{fig:dGdM_pieta}--\ref{fig:dGdM_kpi}, we illustrate the mass distribution results obtained with the parameters from Table~\ref{tab:param}. It is evident that our theoretical calculations closely replicate the experimental data.

In Fig.~\ref{fig:dGdM_pieta}, we can see a clear peak dominating the $\pi^+\eta$ spectrum around $1.0$ GeV, corresponding to the $a_0(980)$ resonance, which, in our model, is encoded in the $T$-matrix elements $t_{\eta\pi^+,\,\eta\pi^+}$, and $t_{\bar{K}^0 K^+,\, \eta\pi^+}$, defined in Eqs.~\eqref{eq:t1} and \eqref{eq:t2}, respectively. In contrast, both the $\bar{K}^0\pi^+$ scattering term in Eq.~\eqref{eq:t3} and the contributions from $K_0^*(1430)$ have relatively small strengths compared to the $a_0(980)$ resonance, particularly around $1.0$ GeV, where their corresponding strengths are 8-9 times smaller. Consequently, we can claim that the peak observed in the experiment can be identified as the $a_0(980)$ state. 
Furthermore, it is crucial to note that due to interference with other contributions, particularly with the tree level in Eqs.~\eqref{eq:tee} and \eqref{eq:tie}, the lineshape of the $a_0(980)$ is broader than those observed in other reactions. Indeed, the apparent width of approximately $150$ MeV exceeds the range of 50-100~MeV as stated in the PDG~\cite{pdg2023}, or the width observed in the $\chi_{c 1} \to \eta \pi^{+} \pi^{-}$ decay~\cite{BESIII:2016tqo} (see also Ref.~\cite{Liang:2016hmr}).

In Fig.~\ref{fig:dGdM_keta}, we reproduce a double hump structure in the $\bar K^0 \eta$ mass distribution.
The $K^*_0(1430)$ spectrum has a peak structure at $M_{\bar K^0 \eta} = M_{K^*_0} = 1425$~MeV with the width $\Gamma_{K^*_0} = 270$~MeV as seen in Eq.~\eqref{eq:K0star}.
The hump structure comes from the interference between the $K^*_0(1430)$ and $a_0(980)$ contributions.
Thus, the phase exp$(i\phi)$ in Table~\ref{tab:param} is required to be negative. 

Fig.~~\ref{fig:dGdM_kpi} shows no distinct peak structure in the $\bar{K}^0\pi^+$ mass distribution, except for a broad bump around $1.25$ GeV. The $\bar{K}^0\pi^+$ spectrum also exhibits 
a relatively large contribution, with a discontinuity at $1.05$ GeV. This behavior 
arises from the cut-off mass $M_{\textrm{cut}}$ applied to the contribution from the 
products $G\cdot t$ in Eqs.~\eqref{eq:tee} and \eqref{eq:tie}, following the prescription discussed 
in Refs.~\cite{Debastiani:2016ayp,Toledo:2020zxj}. The introduction of this cut is essential for extrapolating $Gt$ to high energies, given that the two-body amplitudes from the chiral unitary approach 
are applicable up to about $1200$ MeV. 
Furthermore, it is worth noting that, as discussed in Refs.~\cite{Debastiani:2016ayp,Toledo:2020zxj}, the overall results exhibit minimal dependence on the parameters associated with these cuts. In this study, we adopt $M_{\textrm{cut}} = 1050$ MeV and $\alpha = 0.0037$ MeV$^{-1}$. According to our findings, the impact of $M_{\textrm{cut}}$ is noticeable only in the $\bar{K}^0\pi^+$ distribution. Moreover, when exploring different values for the $M_{\textrm{cut}}$, such as $1150$ MeV as shown in subsection~\ref{subsec}, we observe no significant influence on the distribution lineshapes, except in the $\bar{K}^0\pi^+$ spectrum, where the dip shifts from $1050$ MeV to $1150$ MeV. Thus, we conclude that the dip in our model is not physical, and we could have a smooth curve in that region.  
The relevant thing is that this change in $M_{\rm cut}$ has little repercussion in the $\bar K^0 \eta$ mass distribution and in particular in the $\pi^+ \eta$ distribution, where the effects are negligible. The latter one is the most relevant channel in the present work, where we are concerned about the effect of the $a_0(980)$ in this reaction, and, hence, the conclusions that we draw about the effect of the $a_0(980)$ are not affected by the mentioned uncertainties.

\begin{table}[!tb]
\centering
\caption{\label{tab:param} 
Values of the parameters from the fit. }
\begin{tabular}{cc}
\hline
Parameters & \\
\hline
$\mathcal{C}$ &~~~~$486.90 $ \\
$\mathcal{D}$ &~~~~$63.94 $  \\
$\beta$ &~~~$0.70$  \\
$\phi$ &~~~$-2.16 $ radians   \\
\hline
\end{tabular}
\end{table}

\subsection{Theoretical uncertainties}\label{subsec}
In this subsection, we show the results of several tests to determine some uncertainties of our theoretical models to the three mass distributions. 
%

\subsubsection{Effect of the cut mass $M_{\rm cut}$}
\begin{figure*}[!htb]
\centering
\includegraphics[width=\textwidth]{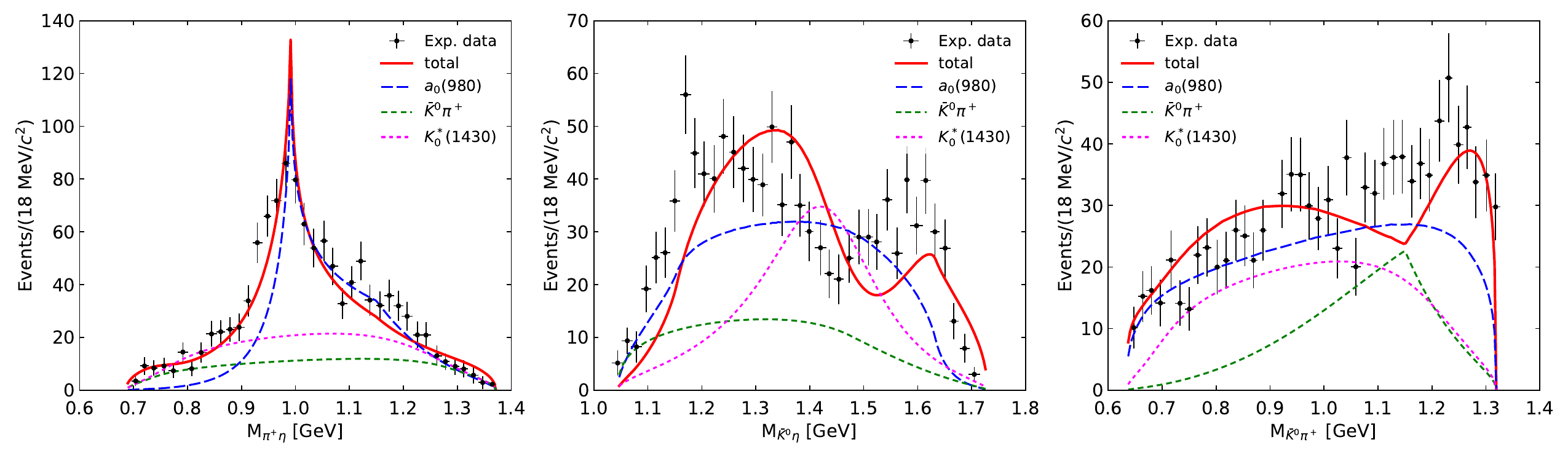}
\vspace{-0.5cm}
\caption{ 
The mass distributions of $\pi^+ \eta$ (left), $\bar K^0 \eta$ (middle), and $\bar K^0 \pi^+$ (right) with fixed $M_{\rm cut}=1150$~MeV. The parameters $\mathcal{C}= 532.04$, $\mathcal{D} = 90.28$, $\beta = 0.70$, and $\phi = -2.07$~radians are used.
} 
\label{fig:dGdM_Mcut1150}
\end{figure*}

 Figure~\ref{fig:dGdM_Mcut1150} shows the results for the considered distributions with $M_{\textrm{cut}}$ fixed at $1150$ MeV. Following a new fit in this setting, we obtain the parameters $\mathcal{C}= 532.04$, $\mathcal{D} = 90.28$, $\beta = 0.70$, and $\phi = -2.07$ radians. As a result, while there are noticeable changes in the strengths of individual contributions, the lineshapes for the first two distributions show only minimal differences, especially in the $\pi^+\eta$ distribution where the $a_0(980)$ peak is evident. Conversely, the dip in the $\bar{K}^0\pi^+$ spectrum has shifted to $1150$~MeV compared to its position in Fig.~\ref{fig:dGdM_kpi}. As discussed earlier, this dip is directly influenced by the parameter $M_{\textrm{cut}}$. Nevertheless, we observe only slight changes in the strength and shape of the $\bar{K}^0\pi^+$ distribution.


\subsubsection{Effect of the parameter $\beta$}
\begin{figure*}[!htb]
\centering
\includegraphics[width=\textwidth]{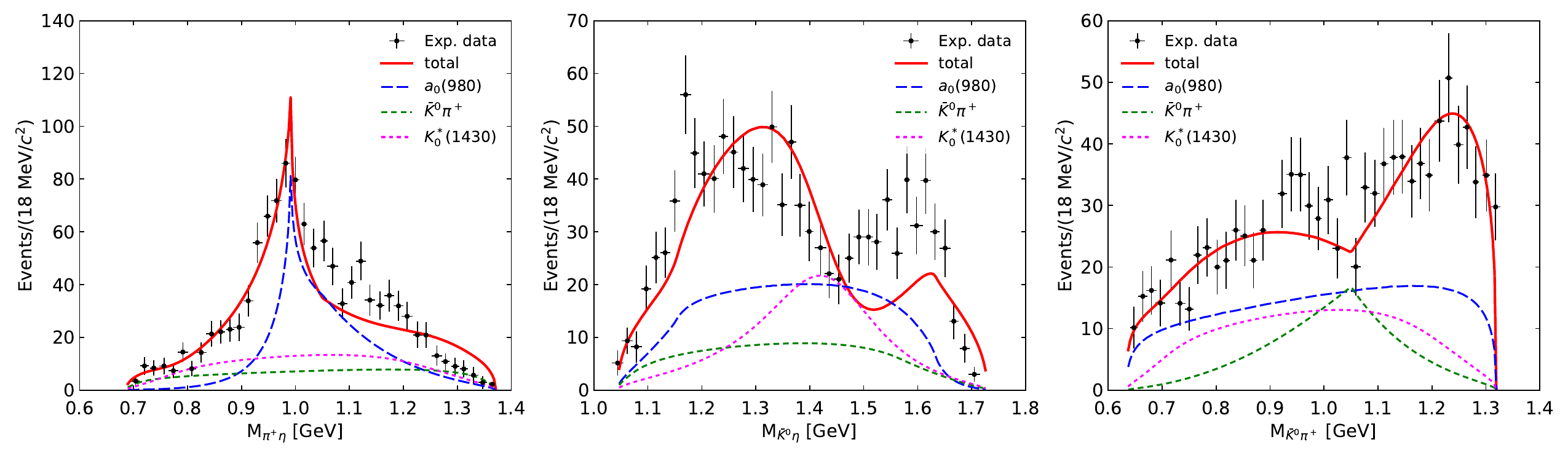}
\vspace{-0.5cm}
\caption{
The mass distributions of $\pi^+ \eta$ (left), $\bar K^0 \eta$ (middle), and $\bar K^0 \pi^+$ (right) with fixed $M_{\rm cut}=1050$~MeV. The parameters $\mathcal{C}= 691.80$, $\mathcal{D} = 71.29$, $\beta = 0.33$, and $\phi = -2.29$~radians are used. }
\label{fig:dGdM_beta0.33}
\end{figure*}

The parameter $\beta$ gives the relative weight of the internal emission mechanism to the external emission. 
The value of the parameter $\beta$ might be expected to be of the order of $1/N_c$. Here we restrict the value of the $\beta$ within $[-0.33:0.33]$ and we make a fit. In Fig.~\ref{fig:dGdM_beta0.33}, we show the three mass distributions with $\beta=0.33$ which is obtained from the best fit.
The parameters obtained from the fit are $\mathcal{C}= 691.80$, $\mathcal{D} = 71.29$, $\beta = 0.33$, and $\phi = -2.29$~radians.
Once again, we see that the changes are not significant.

\subsubsection{Effect of the $K_0^*(1430)$ mass}
\begin{figure*}[!htb]
\centering
\includegraphics[width=\textwidth]{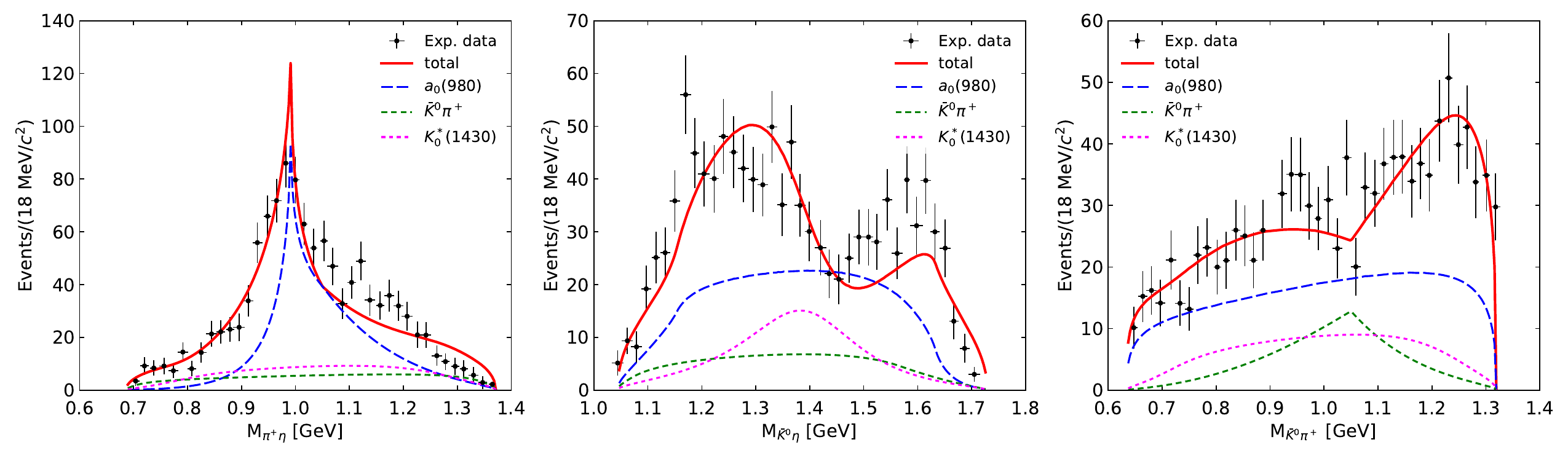}
 \vspace{-0.5cm}
\caption{ 
The mass distributions of $\pi^+ \eta$ (left), $\bar K^0 \eta$ (middle), and $\bar K^0 \pi^+$ (right) with fixed $M_{K^*_0} = 1385$~MeV and fixed $M_{\rm cut}=1050$~MeV. The parameters $\mathcal{C}= 473.34$, $\mathcal{D} = 57.27$, $\beta = 0.70$, and $\phi = -2.39$~radians are used. }
\label{fig:dGdM_K0star}
\end{figure*}

The mass of the $K_0^*(1430)$ has a relatively large uncertainty, $M_{K^*_0} = 1425 \pm 50$~MeV~\cite{pdg2023}. Here we take $M_{K^*_0} = 1385$~MeV, which is consistent within the error bars, and then we perform the same calculations for the three invariant mass distributions. The parameters obtained from the fit are $\mathcal{C}= 473.34$, $\mathcal{D} = 57.27$, $\beta = 0.70$, and $\phi = -2.39$~radians.
We show the three mass distribution in Fig.~\ref{fig:dGdM_K0star}.
The resulting calculation of the $\bar K^0 \pi^+$ mass distribution is in better agreement with the data because the peak position of $K^*_0(1430)$ has moved a bit to the left from Fig.~\ref{fig:dGdM_keta}.

In conclusion, we can see the clear peak of the $a_0(980)$ contribution even considering the uncertainties.
We found that the $D^+ \to \bar K^0 \pi^+ \eta$ reaction is a good reaction to see the peak structure of  $a_0(980)$ in the $\pi^+ \eta$ mass distribution.
Yet, we also explained why the observed shape and width do not correspond to those seen in other experiments, through the interference with other terms that appear necessarily linked to the $a_0(980)$ production in our theoretical approach.

\section{Conclusions}
   We have performed an analysis of the $D^+ \to \bar K^0 \pi^+ \eta$ reaction based on the picture of the $a_0(980)$ resonance as dynamically generated from the interaction of the $\pi \eta$, $K \bar K$ channels, which emerges from the study of the chiral unitary approach and has been tested in many previous reactions. We show that this reaction is drastically different from the apparently analogous one $D^0 \to K^- \pi^+ \eta$,  and we trace it to the absence of a $K^*$ contribution in the $D^+ \to \bar K^0 \pi^+ \eta$ reaction, while it is the driving term in the $D^0 \to K^- \pi^+ \eta$ one. This leads to a much cleaner signal of the $a_0(980)$ excitation in the 
$D^+ \to \bar K^0 \pi^+ \eta$ reaction as seen in the experiment. 
 
    In our study, we begin by looking at the dominant weak decay modes of external and internal emission at the quark level. Then we proceed with the hadronization of the $q \bar q$ pairs into two mesons and finally we allow the meson pairs to interact. We also take into account the contribution of the  $\bar K^*_0(1430)$ as done in the experimental analysis and fit it to the data. Up to this empirical information and a global normalization constant, our framework only depends on the relative strength of the internal emission mechanism to the external emission one, which is also a fit parameter, and we get a result barely consistent with the expected large $N_c$ reduction.  With this framework, we obtain a fair reproduction of the three mass distributions and we observe that the driving term in this reaction is the excitation of the $a_0(980)$. Yet, we note that the shape obtained has a much larger width than the one observed in other experiments, and we trace back this feature to the interference of the dominant $a_0(980)$ excitation with the other mechanisms that are also generated in the final state interaction of the mesons produced, together with the $\bar K^*_0(1430)$ excitation. This explanation is important to prevent taking the present shape of the $a_0(980)$ as a measure of the width of the $a_0(980)$ state, and reconcile the findings of the $D^+ \to \bar K^0 \pi^+ \eta$ reaction with the shapes of the $a_0(980)$ seen in other experiments.

\begin{acknowledgments}
One of us, W.H.L would like to thank Prof. Bai-Cian Ke and Prof. Liao-Yuan Dong for helpful discussion.
N.~I. and J.~M.~D. would like to express gratitude to Guangxi Normal University for their warm hospitality, as part of this work was conducted there.
This work is partly supported by the National Natural Science Foundation of China (NSFC) under Grants No. 11975083 and No. 12365019, and by the Central Government Guidance Funds for Local Scientific and Technological Development, China (No. Guike ZY22096024).
J. M. Dias acknowledges the support from the Chinese Academy of Sciences under Grants No. XDB34030000 and No. YSBR-101; by the National Key R\&D Program of China under Grant No. 2023YFA1606703. 
This project has received funding from the European Union Horizon 2020 research and innovation
programme under the program H2020-INFRAIA-2018-1, grant agreement No. 824093 of the STRONG-2020 project.
\end{acknowledgments}


\begin{thebibliography}{}
\bibitem{BESIII:2020pxp}
M.~Ablikim \textit{et al.} [BESIII],
Phys. Rev. Lett. \textbf{124}, 241803 (2020).

\bibitem{BESIII:2023htx}
M.~Ablikim \textit{et al.} [BESIII],
[arXiv:2309.05760 [hep-ex]].

\bibitem{Belle:2020fbd}
Y.~Q.~Chen \textit{et al.} [Belle],
Phys. Rev. D \textbf{102}, 012002 (2020).

\bibitem{Toledo:2020zxj}
G.~Toledo, N.~Ikeno and E.~Oset,
Eur. Phys. J. C \textbf{81}, 268 (2021).


\bibitem{Xie:2014tma}
J.~J.~Xie, L.~R.~Dai and E.~Oset,
Phys. Lett. B \textbf{742}, 363-369 (2015).


\bibitem{Oller:1997ti}
J.~A.~Oller and E.~Oset,
Nucl. Phys. A \textbf{620}, 438-456 (1997);
[erratum: Nucl. Phys. A \textbf{652}, 407-409 (1999)]

\bibitem{Kaiser:1998fi}
N.~Kaiser,
Eur. Phys. J. A \textbf{3}, 307-309 (1998).

\bibitem{Markushin:2000fa}
V.~E.~Markushin,
Eur. Phys. J. A \textbf{8}, 389-399 (2000).

\bibitem{Nieves:1998hp}
J.~Nieves and E.~Ruiz Arriola,
Phys. Lett. B \textbf{455}, 30-38 (1999).


\bibitem{Chau:1982da}
L.~L.~Chau,
Phys. Rept. \textbf{95}, 1-94 (1983).


\bibitem{Bramon:1992kr}
A.~Bramon, A.~Grau and G.~Pancheri,
Phys. Lett. B \textbf{283}, 416-420 (1992).

\bibitem{Ikeno:2021kzf}
N.~Ikeno, M.~Bayar and E.~Oset,
Eur. Phys. J. C \textbf{81}, 377 (2021).

\bibitem{pdg2023}
R.~L.~Workman \textit{et al.} [Particle Data Group],
PTEP \textbf{2022}, 083C01 (2022).

\bibitem{Lin:2021isc}
J.~X.~Lin, J.~T.~Li, S.~J.~Jiang, W.~H.~Liang and E.~Oset,
Eur. Phys. J. C \textbf{81}, 1017 (2021).

\bibitem{BESIII:2016tqo}
M.~Ablikim \textit{et al.} [BESIII],
Phys. Rev. D \textbf{95}, 032002 (2017).

\bibitem{Liang:2016hmr}
W.~H.~Liang, J.~J.~Xie and E.~Oset,
Eur. Phys. J. C \textbf{76}, 700 (2016).

\bibitem{Debastiani:2016ayp}
V.~R.~Debastiani, W.~H.~Liang, J.~J.~Xie and E.~Oset,
Phys. Lett. B \textbf{766}, 59-64 (2017).


\end{thebibliography}

\end{document}